\documentclass[10pt]{iopart}
\usepackage{amssymb,graphicx}  

\expandafter\let\csname equation*\endcsname\relax
\expandafter\let\csname endequation*\endcsname\relax
\usepackage{float}
\usepackage{bm}
\usepackage{bbold}
\usepackage[colorlinks=true, linkcolor=blue]{hyperref}
\usepackage{color, soul}
\usepackage{xcolor}
\usepackage[utf8]{inputenc}
\DeclareUnicodeCharacter{2212}{-}
\DeclareUnicodeCharacter{B0}{\textdegree}
\usepackage{chemformula}
\DeclareUnicodeCharacter{0301}{\'{e}}
\usepackage{gensymb}
\usepackage[T1]{fontenc}

\usepackage{float}

\begin{document}

\title[]{Driving skyrmions with low threshold current density in  Pt/CoFeB thin film}

\author {Brindaban Ojha$^{1,+}$, Sougata Mallick$^{2,+}$, Sujit Kumar Panigrahy$^2$, Minaxi Sharma$^1$, Andr\'e Thiaville$^2$, Stanislas Rohart$^{2,*}$, Subhankar Bedanta$^{2,*}$}

\address{$^1$Laboratory for Nanomagnetism and Magnetic Materials (LNMM), School of Physical Sciences, National Institute of Science Education and Research (NISER),  An OCC of Homi Bhabha National Institute (HBNI), Jatni 752050, Odisha, India}

\address{$^2$Laboratoire de Physique des Solides, Universit\'e Paris-Saclay, CNRS UMR 8502, F-91405 Orsay Cedex, France}

\address{$^+$These authors contributed equally to this work.}

\address{$^*$Corresponding author.}
\address{E-mail address: stanislas.rohart@universite-paris-saclay.fr, sbedanta@niser.ac.in}

\vspace{10pt}
\date{\today}

\begin{abstract}
Magnetic skyrmions are topologically stable spin swirling particle like entities which are appealing for next generation spintronic devices. The expected low critical current density for the motion of skyrmions makes them potential candidates for future energy efficient electronic devices. Several heavy metal/ferromagnetic (HM/FM) systems have been explored in the past decade to achieve faster skyrmion velocity at low current densities. In this context, we have studied Pt/CoFeB/MgO heterostructures in which skyrmions have been stabilized at room temperature (RT). It has been observed that the shape of the skyrmions are perturbed even by the small stray field arising from low moment magnetic tips while performing the magnetic force microscopy (MFM), indicating presence of low pinning landscape in the samples. This hypothesis is indeed confirmed by the low threshold current density to drive the skyrmions in our sample, at velocities of few $\sim$10m/s. 

\end{abstract}

%
\vspace{2pc}

%
%
\ioptwocol

\section{Introduction}

Since the proposition of employing skyrmions (chiral textures obtained in material with broken inversion symmetry) in spintronic devices \cite{fert2017magnetic}, a collective research effort has addressed the key aspects of this subject. Skyrmions are topologically protected since the spin configuration cannot be twisted continuously to another magnetic state with a different topological number. \cite{fert2017magnetic} Further, their solitonic nature  allows them to behave like particles under the influence of the electrical excitations. These properties make them promising candidates for logic and storage technology. \cite{nagaosa2013topological,fert2017magnetic}. In the experimental perspective, there are three major challenges: (i) stabilization of skyrmions at room temperature, (ii) deterministic nucleation of skyrmions, and (iii) efficient motion of skyrmions under spin Hall effect (SHE). Over the last decade, many experimental works \cite{sampaio2013nucleation,soumyanarayanan2017tunable,jiang2015blowing,woo2016observation}  have been focused in the aforementioned directions to achieve the ambitious goal of skyrmion based device applications. However, controlled nucleation and motion of individual skyrmions in nanotracks with low power consumption still remains a challenge.
\\The topological states can emerge due to spin frustration in single-crystal or amorphous materials with large coordination number.\cite{zhang2017skyrmion,streubel2021chiral} However, in the non-frustrated ordered ferromagnetic materials, the neighboring spins principally exhibit collinear ordering due to Heisenberg exchange interaction. Nevertheless, the presence of large spin-orbit coupling and Dzyaloshinskii-Moriya interaction (DMI) due to broken symmetry can lead to stabilization of non-collinear spin textures viz. skyrmions, \cite{soumyanarayanan2017tunable,jiang2015blowing,woo2016observation,rohart2013skyrmion} since DMI lowers the texture energy and favors chirality. The most widely used combination of such system is a heterostructure of heavy metal (HM)/ferromagnet (FM)/oxide (O).\cite{jiang2015blowing,woo2016observation} For spintronics, this stucture also enables spin-obit torque (SOT) to efficiently manipulate the texture.\cite{sampaio2013nucleation,thiaville2012dynamics}   Recent works have revealed that the skyrmions can be stabilized, nucleated, and driven using external magnetic field \cite{jiang2015blowing,woo2016observation},  electrical field \cite{ma2018electric}, spin polarized current pulses \cite{sampaio2013nucleation,jiang2015blowing,woo2016observation,hrabec2017current}, local field gradient \cite{casiraghi2019individual}, anisotropy gradient \cite{tomasello2018chiral}, etc. Nevertheless, a crucial criterion for device applications is  to drive the skyrmions at low power. This makes materials with low pinning energy landscape indispensable. Different mechanisms (i.e. particle model based numerical analysis, current driven motion of skyrmion via nanotack, skyrmion defect interactions etc. ) suggest that a skyrmion can move around a defect as well as get pinned depending on the pinning sites, applied current density, etc. \cite{sampaio2013nucleation,reichhardt2015collective,lin2013particle} Further, it is well established that at lower current densities, the effect of pinning on magnetic textures is more relevant than at larger current densities. \cite{hanneken2016pinning} One way of optimizing material properties to minimize pinning is to use disordered systems.\cite{kim2017current} However, in spite of several works in this field, the required threshold current density to drive the skyrmions is yet to reach the desired limit for  real-life applications.
\\In this context, we chose the combination of Pt/Co$_{40}$Fe$_{40}$B$_{20}$/MgO to investigate the current driven dynamics of the skyrmions under the influence of SOT. CoFeB has been selected  due to presence of  lower pinning potential in comparison to polycrystalline FM layers (viz., Co, Fe, etc.).\cite {woo2016observation} In addition, it has been observed that a CoFeB-based perpendicular magnetic anisotropic system \cite{fukami2011current} exhibits requirement of a lower threshold current density in current-induced Domain wall motion than a Co/Pt\cite{moore2008high} or Co/Ni \cite{fukami2009relation} system.  Brillouin light scattering (BLS) measurements  have been performed to quantify the DMI in the samples. We show that the threshold current density to drive the skyrmions is significantly lower than the existing literature.

\section{Experimental details}

We have prepared Ta (5 nm)/Pt (6 nm)/ Co$_{40}$Fe$_{40}$B$_{20}$ ($t_{CoFeB}$)/MgO (2 nm)/Ta (3 nm) heterostructure on thermally oxidized Si/SiO$_{2}$(100 nm) substrates. The schematic of the sample structure is shown in Fig. 1(a). We choose SiO$_{2}$ 100 nm since it leads to better signal in the optical measurements \cite{hrabec2017making}. A Ta seed layer has been chosen to promote the (111) growth of Pt as well as to reduce the strain between Pt and the substrate. We also use a 3 nm Ta on the top as a capping layer. The samples are named as S1, S2 for $t_{CoFeB}$ = 1.5 and 1.6 nm, respectively. However, $t_{CoFeB}$ has been scanned from 1.1 to 1.7 nm to find the spin reorientation transition (SRT) and then we kept the $t_{CoFeB}$ values close  to the SRT to balance the total energy of the system favorable to host skyrmions. The sample preparation was performed in a high-vacuum chamber consisting of sputtering and e-beam evaporators. The base pressure of the chamber was better than $8\times10^{-8}$ mbar. Ta, Pt, and CoFeB layers were deposited using DC magnetron sputtering while e-beam evaporation technique was employed to prepare MgO. The substrate has been rotated with 10 rpm during deposition to improve uniformity of the layers. Further, for deposition of CoFeB the Ar flow was kept at 10 sccm which allows uniform growth of the thin film. The growth pressure and rate of deposition for CoFeB were $8\times10^{-4}$ mbar and 0.01 nm/s, respectively. We have used a commercially purchased Co$_{40}$Fe$_{40}$B$_{20}$ stoichiometric target to prepare our thin films. However, from the Energy Dispersive  X-Ray (EDX) analysis measurements (data shown in supplementary information, Fig. S1) for sample S2, it is observed that the amounts of Co, Fe, and B are found to be 43\%, 41\%, and 16\%, respectively. The roughness of the sample is observed to be 0.2 nm from AFM measurements.  In order to promote the interfacial perpendicular magnetic anisotropy (PMA), all the samples were annealed $in-situ$ at $600^{\circ}$C for 1 hour in vacuum ($\sim 1\times10^{-7}$ mbar), after deposition. The hysteresis behaviour of the samples have been performed via the magneto optic Kerr effect based microscopy (MOKE) in polar mode.    To quantify the spontaneous magnetization ($M_s$) of the samples we have performed hysteresis measurements via a superconducting quantum interference device (SQUID) magnetometer. Microfabrication of the tracks and the contacts in the samples has been performed using a two-step electron beam lithography (EBL) and Ar ion etching. We deposited Ti (5 nm) and Au (50 nm) contact pads for transport measurements using e-beam evaporation. The imaging of skyrmions and their dynamics under the application of current has been performed by magnetic force microscopy (MFM) using homemade CoCr/Cr tip coating, to minimize tip induced perturbations.\cite{hrabec2017current,chauleau2010} BLS measurements have been performed in Damon-Eshbach geometry  to quantify the interfacial DMI (iDMI) in the samples.

\section{Results and discussion}

The crystalline nature of the samples has been investigated using X-ray diffraction (XRD) measurements. From grazing incidence X-ray diffraction (GIXRD) data, we obtained peaks with low intensity in both the samples at $\sim 45^{\circ}$ and $\sim 68^{\circ}$, which are not present in the virgin substrate (see Fig. S3 in supplementary information). These peaks have been associated to CoFe (110) and (002)-oriented bcc phase previously reported in the literature.\cite{kanak2013x,ikeda2010perpendicular,6971468} The partial crystallization of the CoFeB due to in-situ annealing originates the perpendicular magnetic anisotropy in the system. We observe that the sample with 1.1 nm thick CoFeB leads to PMA with a square hysteresis loop, whereas the sample with $t_{CoFeB}$ =1.7 nm shows hysteresis with in-plane anisotropy (see Fig. S4 in supplementary information). However, the shape of the hysteresis in Fig. 1(b) for samples S1 ($t_{CoFeB}$ =1.5 nm) and S2 ($t_{CoFeB}$ =1.6 nm) show a sharp change in magnetization (from one saturated state) followed by a relative slow variation before reaching the other saturated state. Usually, in the presence of a finite iDMI and vanishing magnetic anisotropy, such hysteresis loop indicates the presence of a textured (stripes, bubbles, skyrmions) phase as the ground state in the sample.  \cite{mallick2022current}. Further, at the part of the hysteresis where the saturation is progressive after the initial drop in magnetization, skyrmion like textures are likely to be stabilized under the presence of a finite external magnetic field. This has been further confirmed by imaging the magnetic states in samples S1 and S2 as shown in Fig. 2.

\begin{figure}[h!]
	\centering
	\includegraphics[width=1\linewidth]{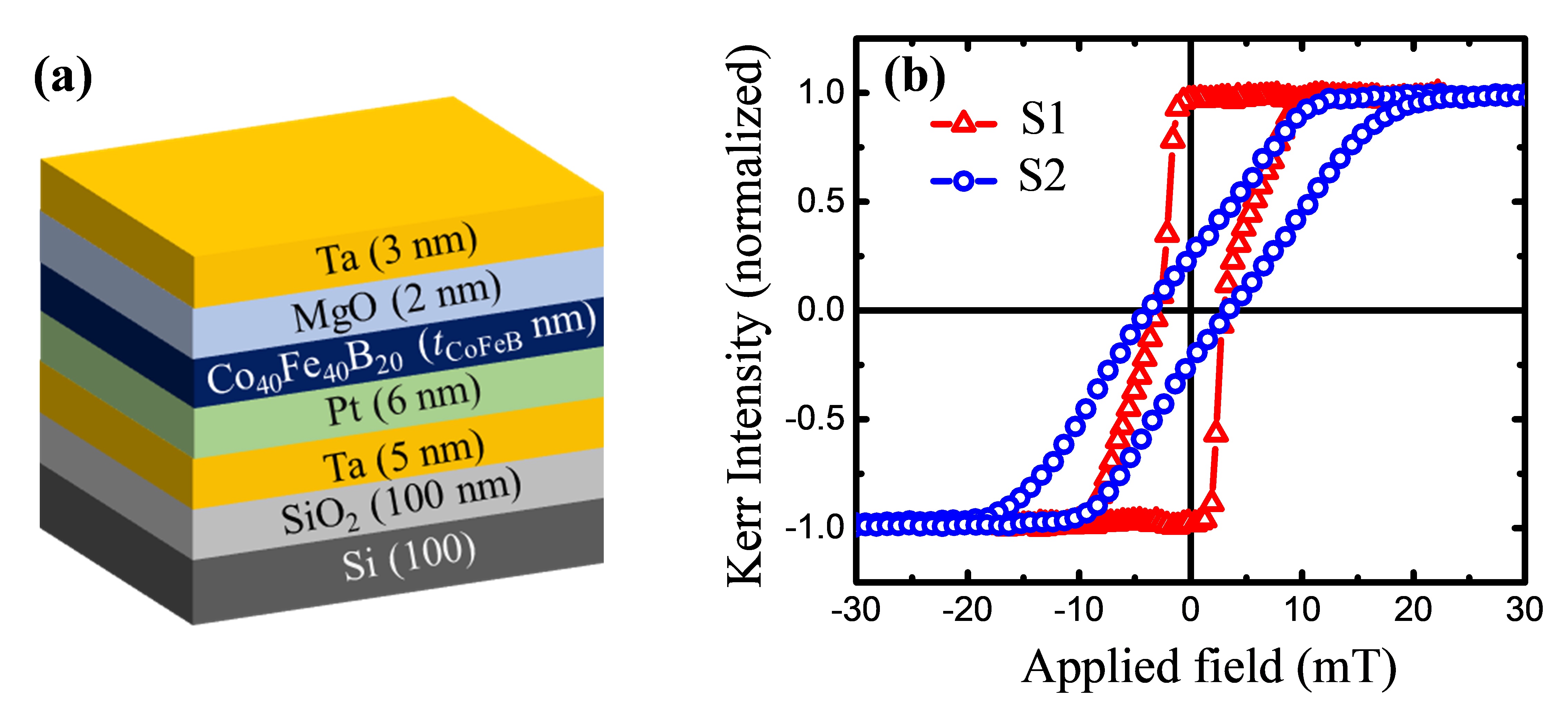}
	\caption{(a) Schematic of sample structure. (b) Hysteresis loop of sample S1 (red) and S2 (blue) in polar MOKE.}
	\label{fig:Fig. 1}
\end{figure}

 We have performed the BLS measurements to quantify the iDMI of the samples. The detailed analysis of calculating the iDMI values from the BLS measurements is shown in the supplementary information. The iDMI constants are $0.33\pm0.03$ mJ/m$^{2}$ and $0.32\pm0.02$ mJ/m$^{2}$ for S1 and S2, respectively, which is similar to previous reports.\cite{tacchi2017interfacial,khan2016effect}

\begin{figure}[h!]
	\centering
	\includegraphics[width=1.0\linewidth]{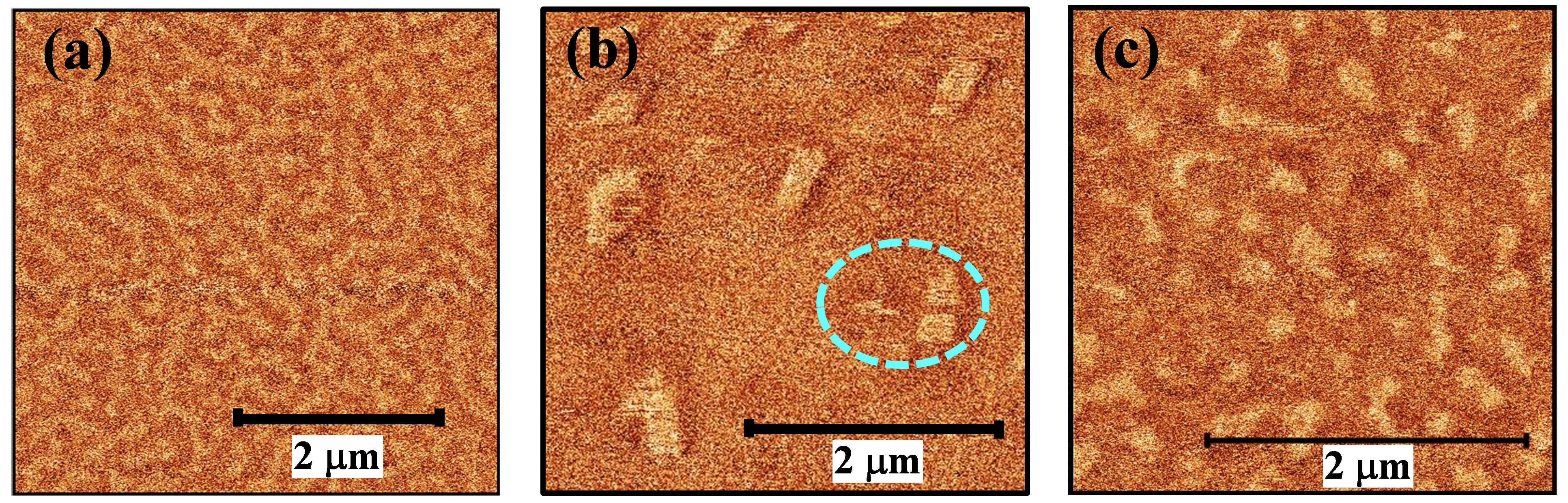}
	\caption{ (a) and (b) MFM images of sample S1 at demagnetized state and at an applied out-of-plane field of 1 mT, respectively. (c) MFM image of sample S2 at an applied field of 3 mT. The cyan circled area in (b) is a guide to the eye for the distorted shape of the skyrmion induced by tip perturbation (horizontal scanning).
	}
	\label{fig:Fig 2}
\end{figure}

In order to stabilize isolated skyrmions, we need to fine tune the domain wall (DW) energy \cite{rohart2013skyrmion}  between two limiting cases: (a) a large positive energy which causes the collapse of the skyrmions, and (b)  a large negative wall energy which destabilizes the collinear order, leading  to isolated skyrmions only at large external magnetic field \cite{hrabec2017current}. One way of achieving this is to keep the FM layer thickness near the SRT to reduce the effective anisotropy energy of the system to near zero. For this work, the thickness of the samples S1 ($t_{CoFeB}$ =1.5 nm) and S2 ($t_{CoFeB}$ =1.6 nm) are chosen close to the SRT. This is confirmed by the presence of skyrmionic states in samples S1 and S2 as observed using MFM. 
Wormlike stripe domains are observed in the demagnetized state indicating that the ground state is the spin spiral state (Fig. 2(a)). By applying OOP fields of 1 and 3 mT (Fig. 2(b) and (c)), isolated skyrmions have been observed in the samples S1, and S2, respectively. The average size of the skyrmions (measured within the accuracy of the MFM) varies in the range of 200-500 nm and 150-300 nm for samples S1 and S2, respectively. We note from fig. 2 (b), and (c) that the shape of the skyrmions is significantly perturbed even by the stray field of the lowest moment magnetic tips. In the area marked by cyan circle in Fig. 2(b), we note that there is a 'flat domain' $\sim$500 nm on the left at the level of the skyrmion which is cut into two domains (skyrmions). This indicates that the skyrmions have jumped to a different pinning point for a few scanning lines. Further, it should be noted that the skyrmions are elongated along the slow scan axis (transverse axis/Y-axis). The reason of such elongation is because the skyrmion has 'more time' to interact with the magnetic tip along the slow scan axis and hence can be dragged progressively providing an elongated shape. Such significant tip induced perturbations indicate the presence of low pinning landscape in the samples. A detailed discussion of the moment of the MFM tips and their effect on the magnetic textures is discussed in Fig. S7 of the supplementary information file. The low pinning landscape of CoFeB can be understood from its structural characteristics which crystallizes into bcc phase when annealed post deposition. It has been reported that the intrinsically amorphous CoFeB crystallizes into bcc structure after annealing beyond $300^{\circ}$C. The presence of an amorphous (and subsequently crystallized bcc phase) maybe expected to lead to a low density structural defects acting as pinning sites in compared to conventional fcc textured 3d ferromagnetic materials. \cite{burrowes2013low,fukami2011current,ikeda2010perpendicular,woo2016observation} Burrows \textit{ et al.}, \cite{burrowes2013low} reported that the depinning field in Ta/CoFeB/MgO films can be as low as 2-3 mT which is an order of magnitude lower than the depinning field for Co/Pt \cite{metaxas2007creep} or Co/Ni \cite{burrowes2010non}. The mean spacing between the consecutive stable pinning sites in CoFeB is ~300 nm  \cite{tetienne2014nanoscale}. This observed low pinning can be corroborated to the low interface roughness, low density of grain boundaries, and better structural coherence in the CoFeB films as compared to other materials hosting skyrmions \cite{zhang2018extrinsic,fukami2011current,ikeda2010perpendicular,tetienne2015nature,jaiswal2017investigation, casiraghi2019individual}.

\begin{figure}[h!]
	\centering
	\includegraphics[width=1.1\linewidth]{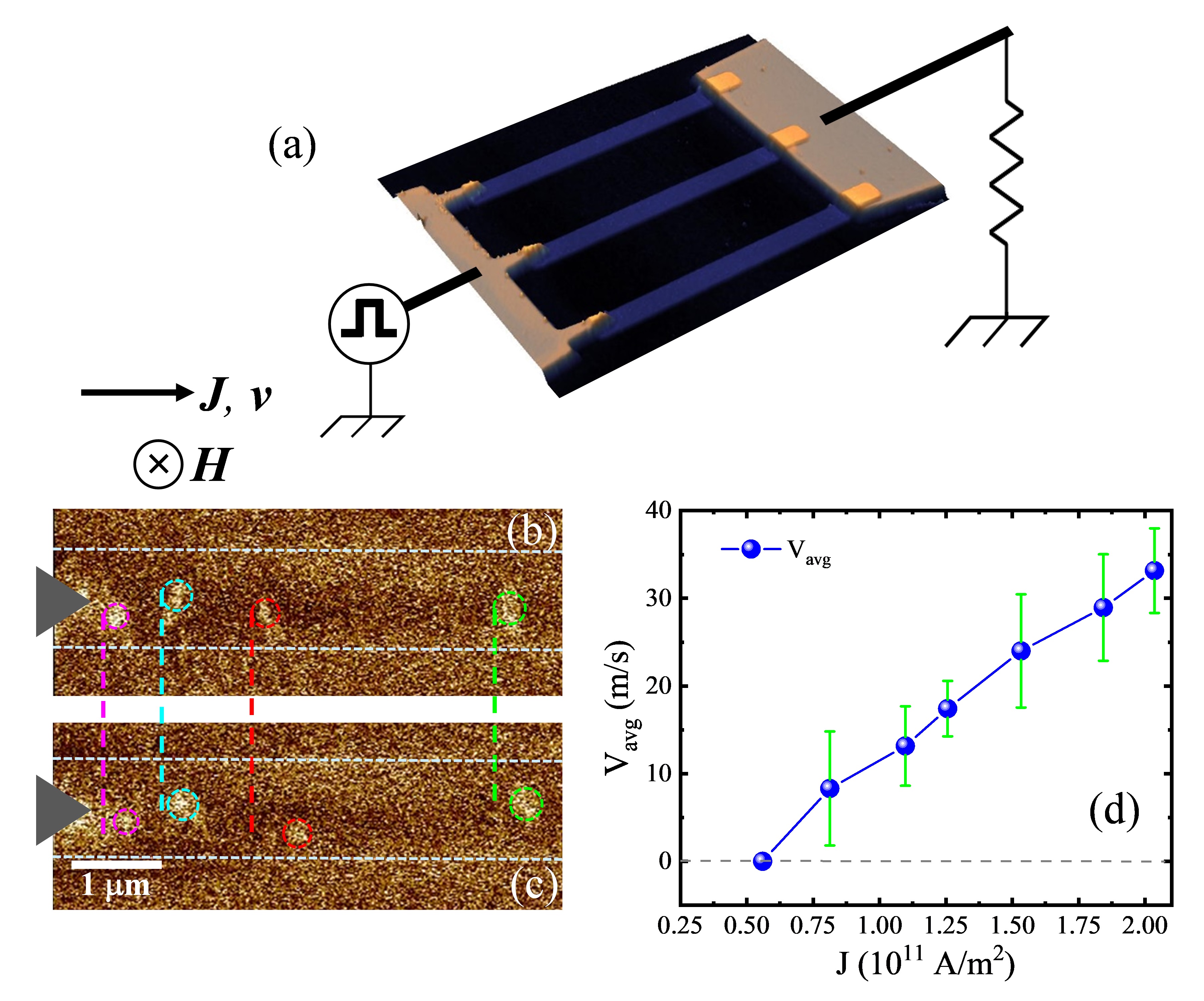}
	\caption{(a) Schematic of the set-up for current induced motion experiment with the AFM image of three 1$\mu$m wide nanotracks (blue) with Ti/Au contacts (light brown) microfabricated on sample S1 (CoFeB: 1.5 nm). (b) and (c) show the position (by coloured circles) and displacement (by coloured lines) of the skyrmions before and after application of one current pulse. The white dashed lines are indicating the boundary of a single nanotrack in the MFM image. Aboves these images, the arrows indicates the current $J$, velocity $v$, and applied field $H$. (d) shows average skyrmion velocity vs current density plot. Blue lines + points are the measured average velocity. The standard deviation is shown by the green error bar. }
	\label{fig:Fig 3}
\end{figure}

Sample S1 is selected for the study of current-induced dynamics due to its lower pinning energy landscape and better control under applied magnetic field.  We start from the demagnetized state and apply a small external field of 1.0 mT  to stabilize the skyrmions in the tracks.  Subsequently, we apply $\sim$ 19 ns current pulses with increasing amplitude. Fig. 3(a) shows the schematic of the sample structure with the AFM images of the nanotracks (3 parallel tracks with width $\sim$1.2 $\mu$m separated by $\sim$2.8 $\mu$m from each other) along with Ti/Au contact pads. The current is injected in the tracks using the point-like contacts on the left and collected at the common ground. We observe that at $J \sim 0.56\times10^{11}$ A/m$^{2}$, the skyrmions in the track do not move and beyond a threshold current density of $\sim 0.8\times10^{11}$ A/m$^{2}$, the skyrmions start moving. The measurements for each current density are repeated in between 3-5 times depending on the number of skyrmions inside the tracks. The velocity of skyrmions is calculated by averaging the movement of these 4-7 number of skyrmions which move due to the pulses. The motion of the skyrmions, opposite to the electron motion direction, confirms the dynamics due to SOT \cite{sampaio2013nucleation,woo2016observation,jiang2015blowing, buttner2017field}.  In order to confirm that the stray field originating from the magnetic tip has minimal/no effect on the motion of the skyrmions, we have checked using \textit{no pulse} applied where the skyrmions were stable (a detailed discussion is given in Fig. S8 of SI).  Fig. 3(b) and (c) show the skyrmion displacements marked by different colours before and after the application of one current pulse. It should be noted that both at lower as well as at higher current densities, the displacements of all the skyrmions are not equal. This is the consequence of the skyrmion hopping due to the presence of defects within the potential landscape \cite{kim2009interdimensional}. The skyrmions advance in the track until they are either pinned or face strong skyrmion-skyrmion repulsion from another pinned skyrmion. We have plotted the average velocity of all the skyrmions present in the track for a particular current density (Fig. 3(d)).    The error bar (green) in Fig. 3(d) corresponds to the standard deviation. The high current velocity data indicate, within the deviation range of the data, that a regime is reached, where defects plays a minor role, analogue to a flow regime \cite{berges2022size}. We obtain skyrmion velocity up-to $\sim$40 m/s for a current density of $\sim 2\times10^{11}$ A/m$^{2}$. Application of higher current densities led to burning of the contacts due to application of large (19 ns) current pulses through the tracks  (see Fig. S9 in supplementary information for details). The threshold current density reported in this work is lower than the ones reported in the literature (see Table 1 in supplementary information) whereas the corresponding velocity ($>$ 10 m/s) is comparable to previous reports.

\section{Conclusions}
In conclusion, we have shown that  skyrmions are stabilized at very low field in Pt/CoFeB/MgO heterostructures at room temperature. We have quantified the iDMI value for our thin films which is similar to earlier reports. We observe significantly lower threshold current densities to drive the skyrmions. Larger skyrmion velocity under shorter current pulses can be expected in similar samples. Here, we have used a single set of optimized deposition conditions (growth pressure, deposition rate, annealing temperature) to prepare the samples. However, the impact of various deposition conditions on pinning potential can be investigated in the future. We believe that the work presented here may be helpful to utilize such low pinning  materials for skyrmionic applications at low power consumption.

\section*{Acknowledgements} 
The authors thank DAE, Govt. of India and the Indo-French collaborative project supported by CEFIPRA (IFC/5808-1/2017), and the French National Research Agency (ANR) (Topsky, ANR-17-CE24-0025)  for providing the research funding. 
The BLS setup was funded by the PhOM and EOE Research Departments of Université Paris-Saclay, CNRS-Institut de Physique, French National Research Agency (ANR) as part of the  ``Investissements d$^{\prime}$Avenir" program (Labex NanoSaclay, ANR-10-LABX-0035) through the BLS@Psay and SPICY projects, and Ile-de-France region through the SESAME IMAGeSPIN (EX039175) project.
We would like to thank Dr. Braj Bhusan Singh for valuable scientific discussion. We would also like to thank Raphael Weil for his help in microfabrication. 

\section*{References}

\bibliographystyle{iopart-num}
\bibliography{reference}

\end{document}



\leavevmode\thispagestyle{empty}\newpage

\section{Sample characterization}
\subsection{Energy Dispersive X-Ray (EDA) analysis measurements }
We have performed the Energy Dispersive X-Ray (EDX) analysis measurements in sample S2 in three different areas to confirm the Co, Fe and B percentage present in the thin films. The amounts of Co, Fe, and B are found to be 43\%, 41\%, and 16\%, respectively (see Fig. S1). So, it can be concluded that in the thin films the composition of CoFeB may be little bit deviated from the actual composition.

\begin{figure}[h!]
	\centering
	\includegraphics[width=1.0\linewidth]{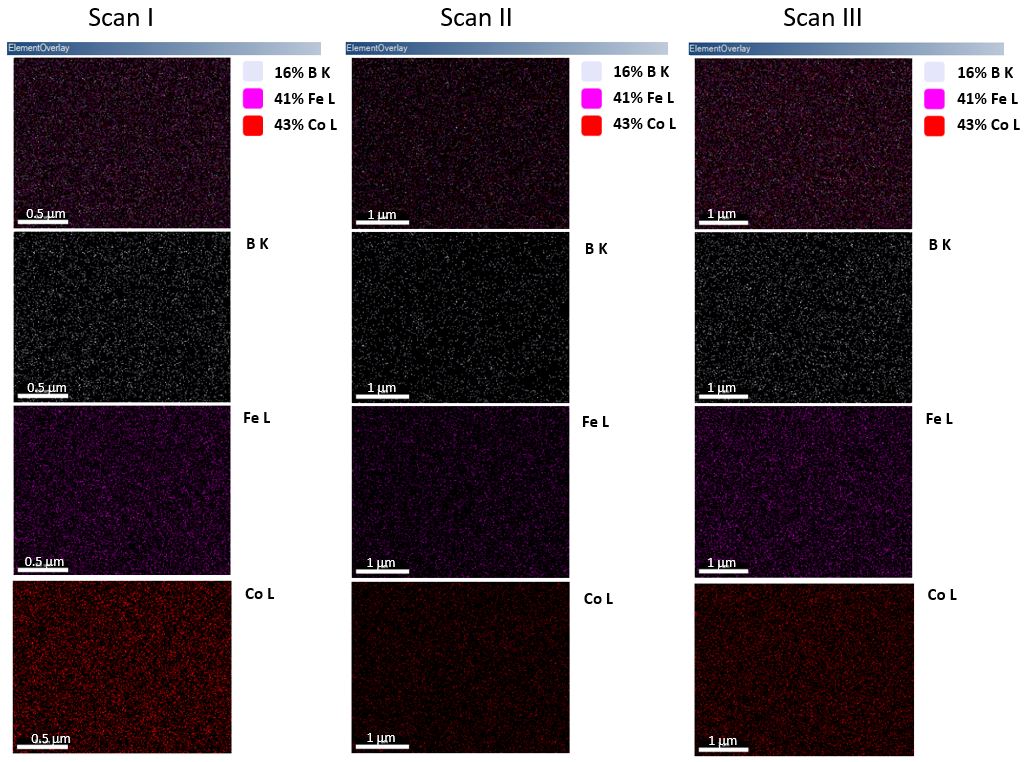}
	\renewcommand{\thefigure}{S\arabic{figure}}
	\caption{ EDX scan on sample S2 at three different areas}
\end{figure}

\subsection{Thickness calibration}
In order to know the exact thickness of each layer of the films, we performed X-ray reflectivity (XRR) measurements. Fig. S2 (a) and (b) show the XRR fits for samples S1 and S2, respectively. From, the fittings it was calculated that the thicknesses of CoFeB are $\sim$ 1.5 and 1.6 nm for the samples S1 and S2, respectively. 
\begin{figure}[h!]
	\centering
	\includegraphics[width=0.9\linewidth]{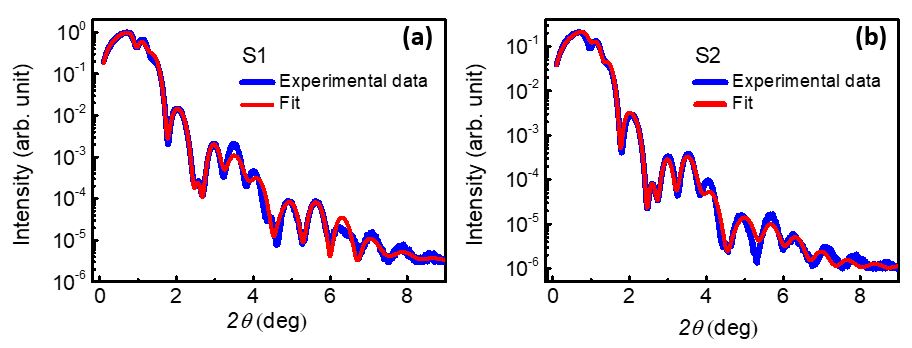}
	\renewcommand{\thefigure}{S\arabic{figure}}
	\caption{ (a) and (b) XRR fits for samples S1 and S2, respectively.}
\end{figure}

\subsection{Structural characterization}

\begin{figure}[h!]
	\centering
	\includegraphics[width=0.5\linewidth]{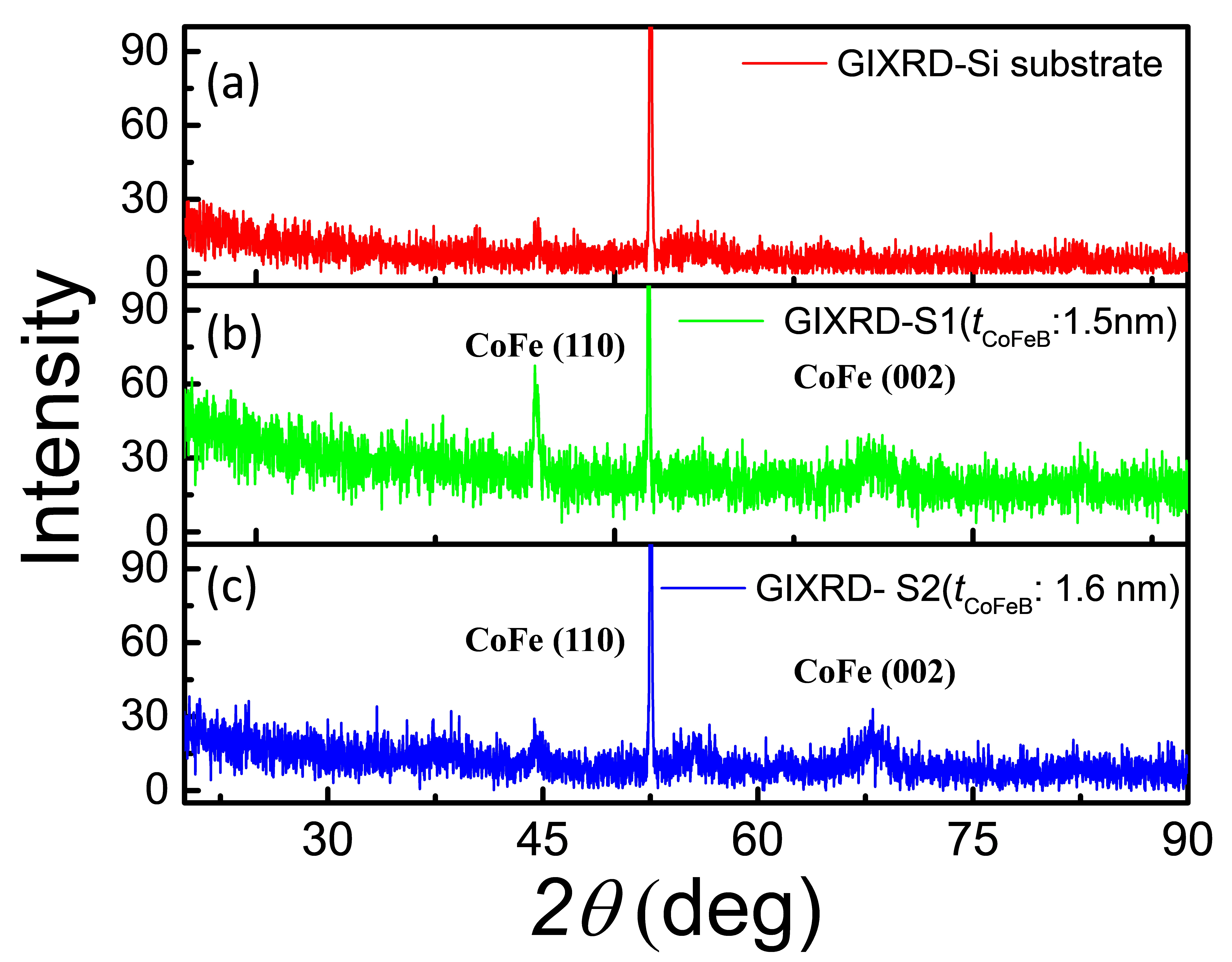}
	\renewcommand{\thefigure}{S\arabic{figure}}
	\caption{ GIXRD measurements of the bare substrate and the samples S1 and S2 are shown in (a), (b) and (c), respectively. For proper comparison, the intensity axis scale is same for all the plots.}
\end{figure}

Fig. S3 shows the grazing incidence X-ray diffraction (GIXRD) data of the samples S1 (CoFeB: 1.5 nm), S2 (CoFeB: 1.6 nm), and the bare substrate. We obtained peaks with low intensity in both the samples at ~ 45° and 68° which have been associated to CoFe bcc phase.\cite{ikeda2010perpendicular,kanak2013x,6971468}
 The rather broad shape of the peaks are due to the low thickness of the thin films. The XRD measurements confirm amorphous to crystalline phase transformation of the CoFeB thin films due to annealing.

\subsection{Magnetometry}

Fig. S4 shows the hysteresis loops of the samples measured by magneto optic Kerr effect based
 (MOKE) microscope. We observe that the sample with 1.1 nm thick CoFeB leads to PMA with a square loop, whereas the sample with $t_{CoFeB}$=1.7 nm shows hysteresis with in-plane anisotropy. 

\begin{figure}[h!]
	\centering
	\includegraphics[width=1.1\linewidth]{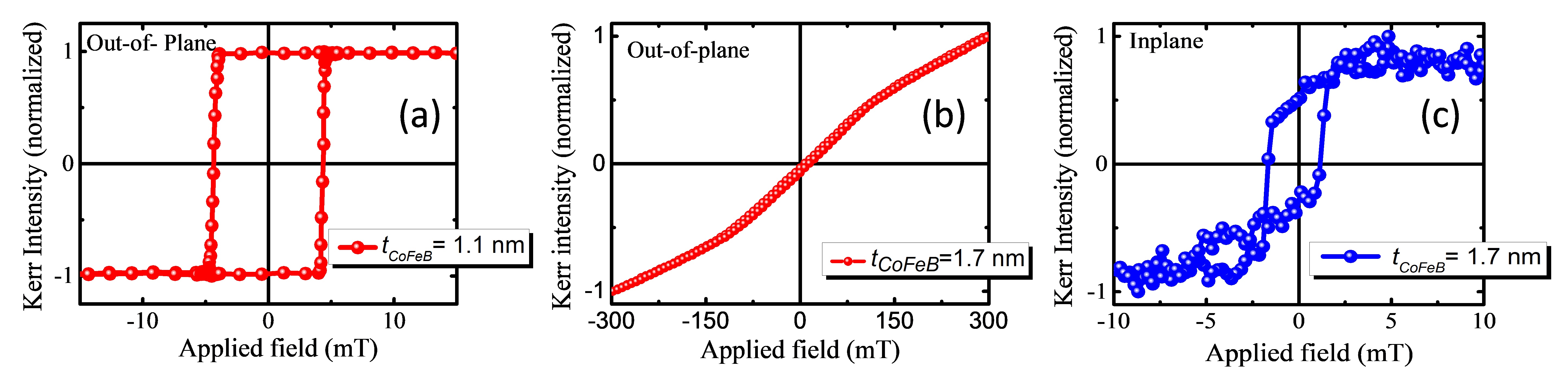}
	\renewcommand{\thefigure}{S\arabic{figure}}
	\caption{ (a) MOKE hysteresis loop for the sample with tCoFeB= 1.1 nm in polar mode.  (b) and (c) Hysteresis loops for the sample with tCoFeB= 1.7 nm measured at room temperature in polar mode and longitudinal modes, respectively.}
\end{figure}


\subsection{iDMI measurement}

Fig. S5 (a) shows the schematic of the measurement geometry using Brillouin light scattering (BLS) in Damon Eshbach (DE) configuration. In this measurement, a horizontally polarized laser beam is incident on a sample surface where the incident photons interact with magnons. The $180^{\circ}$ backscattered photons (arises due to inelastic scattering) are sent through a tandem Fabry-Perot (FP) interferometer to achieve a high spectral resolution and to extract the information about spin wave. Stokes (S) and Anti-stokes (AS) peaks arises due to creation and annihilation of the magnons, respectively, in the BLS spectra and the respective spin wave propagates in the opposite direction. The wave vector of the spin wave is represented as $k_{sw} =  \frac {4\pi sin\theta}{\lambda}$, where {$\theta$} is the incident angle, and {$\lambda $} (=532 nm) is the wavelength of incident laser light. Here, a magnetic field has been applied perpendicular to the incident plane of light. Fig. S2 (b) shows the BLS spectra of the sample S2 measured at  $k_{sw}$= 4.1 $\mu m^{-1}$ and a constant applied field ($\mu_{0} H$ =  371 mT).

\begin{figure}[h!]
	\centering
	\includegraphics[width=0.8\linewidth]{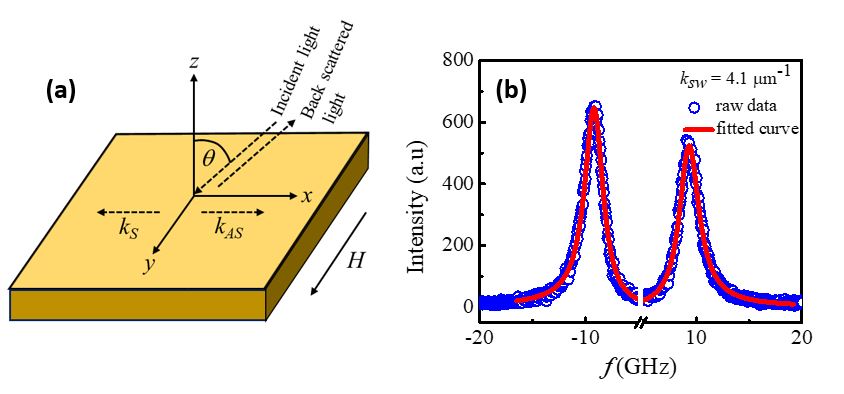}
	\renewcommand{\thefigure}{S\arabic{figure}}
	\caption{ (a) Measurements geometry of BLS technique. (b) BLS spectra of sample S2 measured at a constant applied field ($\mu_{0} H$ =  371 mT).}
\end{figure}

The BLS data have been analyzed using the following spin-wave dispersion relation in DE mode \cite{belmeguenai2015interfacial,tacchi2017interfacial,di2015asymmetric}
\begin{equation}
	f = f_{0}  \pm f_{DMI} = \frac {\gamma\mu_{0}}{2\pi} \sqrt{[H+J {k^{2}_{sw}} + P(k_{sw}) M_{s}][H+J {k^{2}_{sw}} - P(k_{sw}) M_{s} -H_{K_{eff}}]}  \pm \frac {\gamma}{\pi M_{S}} D_{eff} k_{sw}
\end{equation}

where, $\gamma$ is the absolute value of the gyromagnetic factor [$\gamma/2\pi$ = $g \times 13.996$ GHz/T, with $g$ the gyromagnetic factor], $\mu_{0}$ is the vacuum permeability, $J = 2A/(\mu_{0} M_{S}$) is the SW stiffness constant with $A$ the micromagnetic exchange constant, $D_{eff}$ is the effective DMI constant, $H_{K_{eff}}$ is the effective anisotropy, $P(k_{sw})$ = $1 - \frac {1-exp(-|k_{sw}| t_{CoFeB})}{|k_{sw}| t_{CoFeB}}$. The first part of the above equation, $f_{0} = \frac {f_{S} + f_{AS}}{2}$ ($f_{S}$ and $f_{AS}$ peak frequencies of S and AS lines of BLS measurements), represents the average spin wave frequency having no iDMI at the HM/FM interface whereas the second part arises due to iDMI. $M_S$ is the saturation magnetization. The $M_S$ value of sample S1 and S2 is obtained to be $\sim 1.24\times10^{6}$ A/m and $\sim 1.27\times10^{6}$ A/m, respectively.

\begin{figure}[h!]
	\centering
	\includegraphics[width=0.6\linewidth]{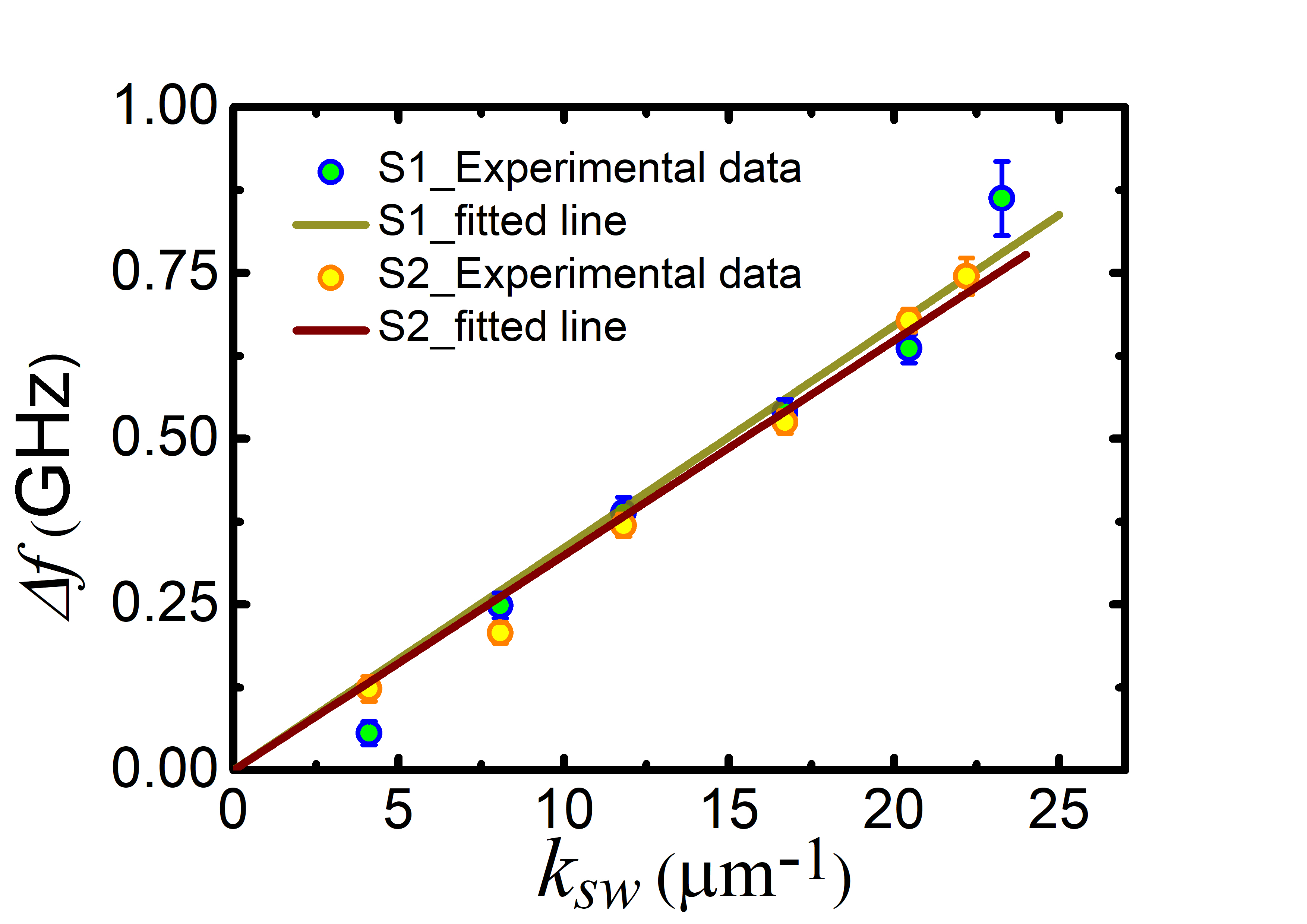}
	\renewcommand{\thefigure}{S\arabic{figure}}
	\caption{ Plot of frequency difference ($\Delta f$) vs spin wave vector ($k_{sw}$) at $\mu_{0} H$ =  371 mT}.
\end{figure}

From equation (1) the frequency difference between S and AS peaks can be written as:
\begin{equation}	
	\Delta f = f_{S} - f_{AS} = \frac {2 \gamma}{\pi M_{S}} D_{eff} k_{sw}
\end{equation}
$\Delta f$ as a function of $k_{sw}$ has been plotted in Fig. S6 for both the samples. From the slope of linear fit, DMI constant $D_{eff}$  can be deduced. The magnitude of iDMI constants are $0.33\pm0.03$ mJ/m$^{2}$ and $0.32\pm0.02$ mJ/m$^{2}$ for S1 and S2,  respectively.

\section{Magnetic force microscopy imaging}

\begin{figure}[h!]
	\centering
	\includegraphics[width=0.6\linewidth]{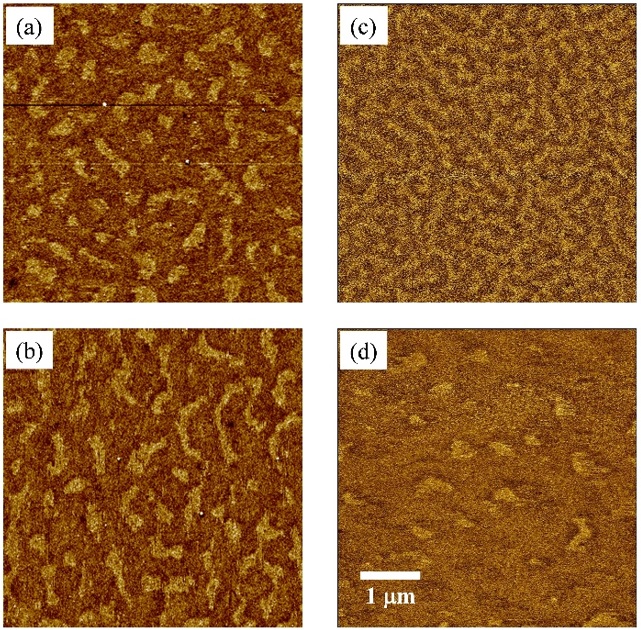}
	\renewcommand{\thefigure}{S\arabic{figure}}
	\caption{ MFM image of the CoFeB sample in the demagnetized state (zero field) using a 15 nm/20 nm  (a-b) and a 5 nm/20 nm (same tip as in the manuscript) (c) CoCr/Cr bilayers tip. Images (a) and (b) show significant pertubration whereas image (c) show the expected stripe phase. Note that the different scan direction (respectively $x$ and $y$ fast scan axis) between images (a) and (b) highlight the deformations due to the magnetic tips that show magnetic textures elongated along the slow scan axis. (d) Image obtained with a 3~mT applied field of the skyrmion state, using the 5 nm/20 nm (c) CoCr/Cr bilayers tip. This image shows that in the skyrmion phase, perturbation cannot be avoided. Thinner magnetic coating prevents  a good signal being obtained from the samples. The phase scale for all three images are the same (300 mDeg). The image scales are identical, with a scale bar shown in (d). }
\end{figure}

The MFM images were realized using two different microscope, a Brucker D3100 and a Park NX10 using two-pass scan, one for the topography and one at a higher distance (20 nm) to get the magnetic signal from the phase of the tip oscillation. In such a measurement, tip induced perturbation should occur during the topographic scan, when the tip to sample distance is the lowest. This is why we have used home-made tips that fix this problem. In addition to a magnetic coating (CoCr), we add a Cr capping layer. This last layer is not only meant to protect the magnetic coating but also increases the magnetic layer to sample distance during the topographic scan. This strategy has been successfully used in several of our studies \cite{chauleau2010,torrejon2012,hrabec2017current} on magnetic solitons (skyrmions and domain walls) and prevent from dramatic and irreversible perturbations.

The consequences of the CoCr layer thickness on the images is shown in Fig. S7. In this figure we compare the same demagnetized state using two tips with CoCr(15 nm)/Cr(20 nm) [images (a) and (b)] and CoCr(5 nm)/Cr(20 nm) [image (c)] (same tip as in the manuscript). The striking difference between the two sets of images is that, contrary to the expected stripe state observed with the thinnest CoCr layer tip, the other tip shows more compact structures. This is the result of a significant perturbation which slices the stripes into more compact domains. Changing the scan direction between the $x$ [image (a)] and $y$ [image (b)] fast scan axis proves the perturbation since the observed textures are systematically elongated along the slow scan axis, probably due to a dragging effect by the magnetic tip. Note that if the thinnest CoCr layer tip produces minute perturbation on the stripe phase, the skyrmion state is still slightly perturbed [image (d) and main figures of the manuscript]. This behaviour is likely to be due to the low pinning landscape of the CoFeB thin films. 
In order to confirm that the tips (CoCr(15 nm)/Cr(20 nm)) used here are soft magnetic in nature we have performed, MFM with the same tips on our previous studied sample \cite{hrabec2017current} of [Pt/Ni/Co/Ni/Au/Ni/Co/Ni/Pt] in which the skyrmions have been imaged without any perturbations (i.e. without any distortion).

\section{Effect of MFM tip on skyrmion motion}

Recently A. Casiraghi \textit{et al.}, have observed that the motion of skyrmions occurs along the direction of the MFM slow scan axis when scanning in two-pass mode \cite{casiraghi2019}. In our work we have also performed the MFM scan in two-pass mode. We have chosen the ``slow scan axis" along the length of the track to further reduce the effect of stray field of the magnetic tip on the motion of the skyrmions. We can see from Fig. S8 that there is no motion of skyrmions observed before and after applying a current pulse of J$\sim$ $0.56\times10^{11}$ A/m$^2$, 19 ns. This clearly shows that there is no motion of skyrmions along the ``slow scan axis" due to tip induced perturbation.

\begin{figure}[h!]
	\centering
	\includegraphics[width=0.5\linewidth]{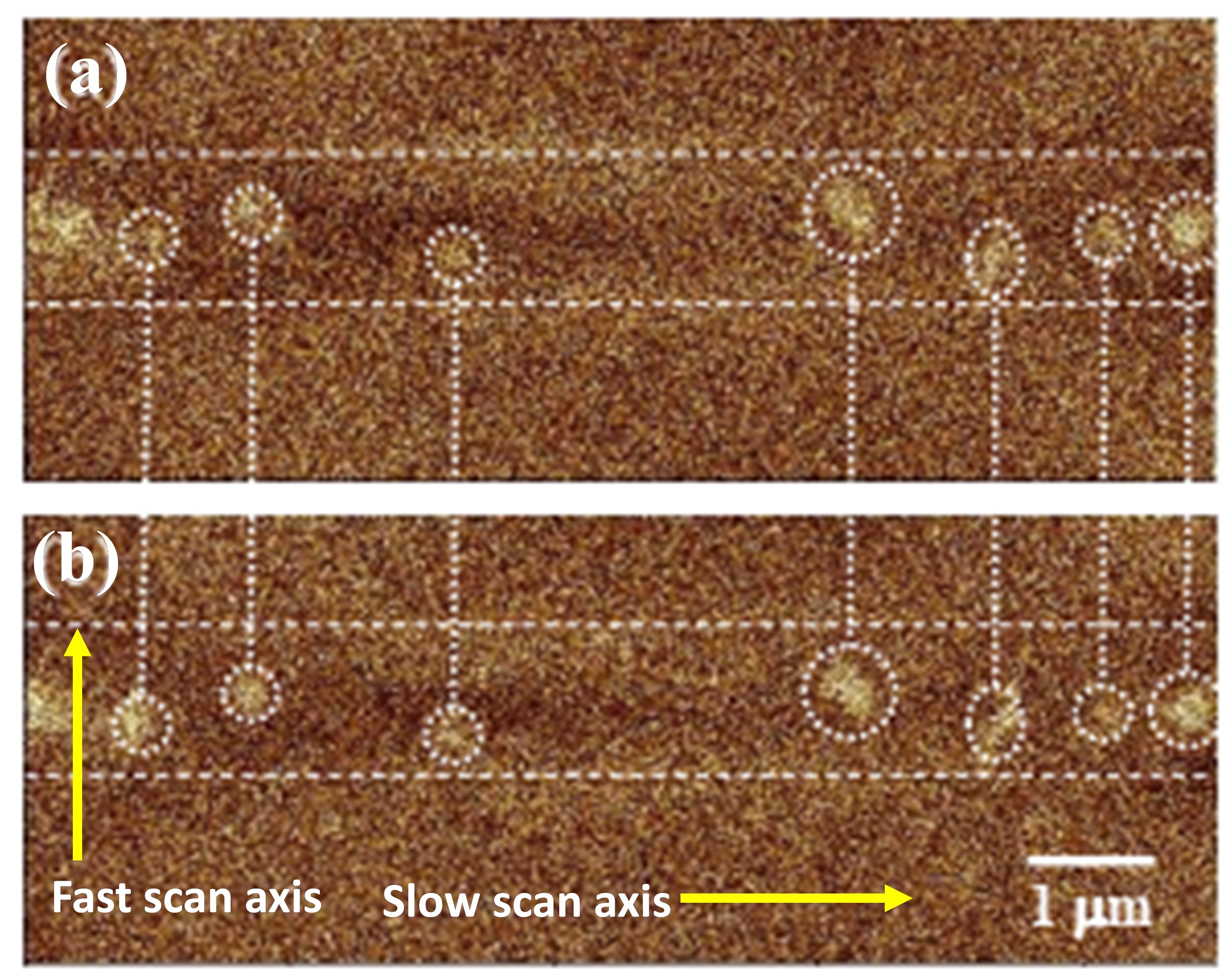}
	\renewcommand{\thefigure}{S\arabic{figure}}
	\caption{(a) and (b) are images of the skyrmions in the nanotrack performed before and after applying a current pulse of J$\sim$ $0.56\times10^{11}$ A/m$^2$ and duration 19 ns. The dashed circles and lines are guide to the eye to represent the position of the skyrmions in the track. This demonstrates that at this current density all the skyrmions in the tracks are pinned and do not move under the applied current pulse.}
\end{figure}

\section{Current pulses}

Current pulses through the sample through a 50~$\Omega$ impedance cable ended by a 50~$\Omega$ load to absorb the pulse power after transmission through the sample as shown in figure S9 (a). Although this system can provide pulses raise time of about 100~ps, the high resistivity substrate significantly degraded the system respond time.

\begin{figure}[h!]
	\centering
	\includegraphics[width=0.9\linewidth]{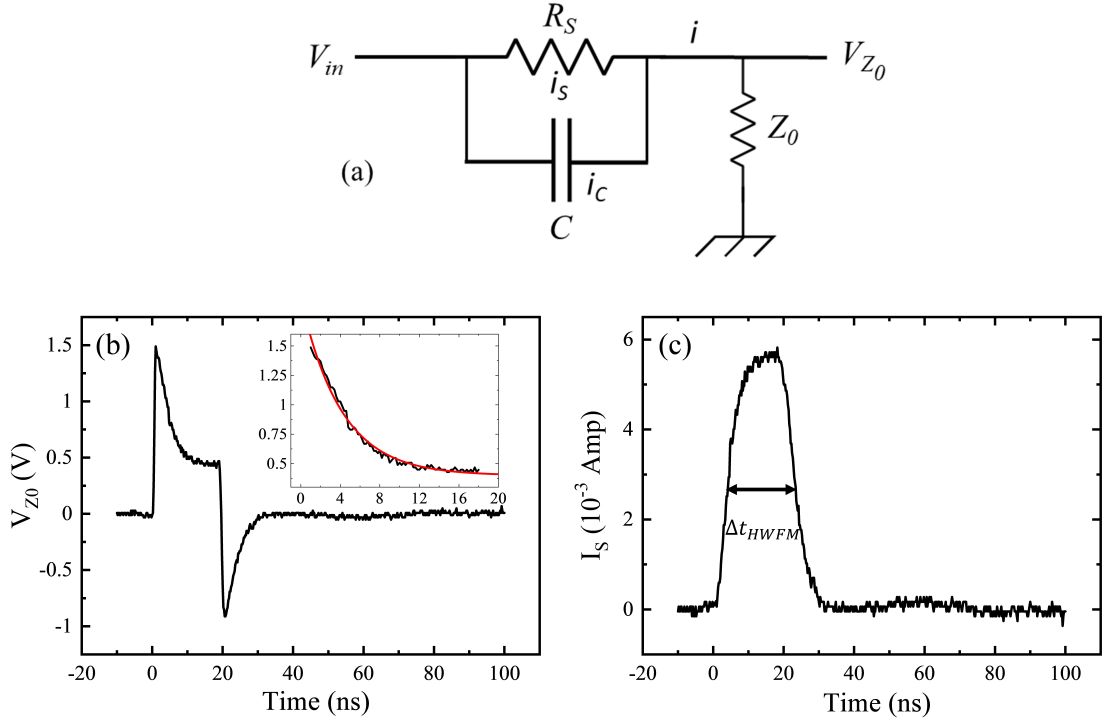}
	\renewcommand{\thefigure}{S\arabic{figure}}
	\caption{(a) Circuit diagram of the sample considering capacitive short through the substrate. (b) Shape of the voltage pulse measured on the 50 $\Omega$ load of an oscilloscope and (c) processed  current pulse flowing through the nanotracks, for a 20 ns long 2 V pulse (note that due to the impedance mismatch at the sample, 25$\%$ of the pulse is reflected before the sample and does not participate to the pulse).}
\end{figure}

The substrate used in the experiment is a Si(001) substrate with a resistivity of 0.01 $\Omega$.cm, covered by 100~nm SiO$_2$. The oxide layer is sufficient to isolate the tracks in DC. However, since the substrate resistance is much lower than the track resistance $R_S$, the current initially flows through the substrate, due to capacitive coupling between the track leads and the substrate, through the SiO$_2$ layer, as sketched in Fig. S9 (a). The pulse measured on the 50 $\Omega$ load of an oscilloscope does not directly correspond to the current through the tracks, but to the sum of the currents through the substrate and the track. At the beginning of the pulse, the current flows directly through the capacitor and the voltage on the load is the one of the pulse. After the transient regime, the current through the capacitor is zero and the voltage on the load is $Z_{0}/(Z_{0}+R_{S})$. From the measurement in Fig. S9 (b), a time constant of 4 ns and a track resistance $R_{S} = 188  \Omega$ are deduced. The current pulse through the sample can be deduced, as shown in Fig. S9 (c). The pulse width is calculated as the full-width at full maxima of the pulse, here 19 ns. The current density through the sample is subsequently calculated by dividing the current with the product of width and thickness of the nanotracks.

\leavevmode\thispagestyle{empty}\newpage

\begin{table*}
	\caption{Comparison of threshold current density with existing literature at room temperature. Two groups of results can be distinguished. In the first one, the measured velocities are well below 1~m/s, while in the second one, the velocities are larger than 10~m/s. In the second group, that concerns the fastest skyrmion, our work displays the smallest threshold current density.} 
	\centering
	\label{tbl:example}
	\begin{tabular}{ p{1.0cm}|p{3cm}|p{5cm}|p{2.8cm}|p{2cm}  }
		Group&Authors& Sample structure &Threshold current density (A/m$^{2}$)&Velocity range (m/s)\\
		\hline
		1    & Jiang $et$ $al.$ \cite{jiang2015blowing}           & Ta/CoFeB/TaO$_{x}$                          & $1.5\times10^{8}$   & $(2.5 - 25)\times10^{-6}$\\
		
		& Yu $et $ $ al.$ \cite{yu2017room}         & Ta/CoFeB/TaO$_{x}$ 
		& $3.5\times10^{9}$   & 0.2 -- 0.4\\
		&Jiang $et $ $ al.$ \cite{jiang2017direct}      & Ta/CoFeB/TaO$_{x}$                   & $6\times10^{9}$     & 0.01 -- 0.75\\
		
		&Tolley $et$ $ al.$ \cite{tolley2018room} & Ta/Pt/Co/Os/Pt/Ta                           & $2.0\times10^{8}$   & $(2.5 - 13)\times10^{-6}$\\
		& & & &\\

		2    &Woo $et $ $ al.$ \cite{woo2016observation}         & [Pt/Co/Ta]$_{15}$ ;   [Pt/CoFeB/MgO]$_{15}$ & $2.0\times10^{11}$  & 1--45; 1--100\\
		
		&Legrand $et $ $ al.$ \cite{legrand2017room}    & Ta/Co/[Pt/Ir/Co]$_{10}$/Pt                  & $2.38\times10^{11}$ & 15 -- 47\\
		
		&Hrabec $et $ $ al.$ \cite{hrabec2017current}      & Pt/FM/Au/FM/Pt, FM= Ni/Co/Ni                & $2.6\times10^{11}$  & 10 -- 60\\
		
		&Woo $et $ $ al.$ \cite{woo2017spin}       & Ta/[Pt/CoFeB/MgO]$_{20}$/Ta                 & $1.5\times10^{11}$  & 5 -- 40\\
		
		&Akhtar $et $ $ al.$ \cite{akhtar2019current} & Ta/Pt/CFA/MgO/Ta                            & $1.6\times10^{11}$  & 10\\
		
		&Juge $et $ $ al.$ \cite{juge2019current} & Ta/Pt/Co/MgO/Ta                             & $3.5\times10^{11}$  & 15 -- 100\\
		
		&Litzius $et $ $ al.$ \cite{litzius2020role}     & [Pt/CoFeB/MgO]$_{15}$                       & $3.2\times10^{11}$  & 1 -- 35\\
		
		&This work                          & Ta/Pt/CoFeB/MgO/Ta                          & $0.8\times10^{11}$  & 9 -- 46\\
	\end{tabular}
\end{table*}


\bibliography{references-supp}